\documentclass[12pt]{iopart}

\usepackage{iopams}
\usepackage[dvips]{graphicx}
\usepackage{epsfig}
\usepackage{latexsym}
\usepackage{color}

\begin{document}

\title[]{Electrostatic trapping and \emph{in situ} detection of Rydberg atoms above chip-based transmission lines}

\author{P. Lancuba and S. D. Hogan}

\address{Department of Physics and Astronomy, University College London, Gower Street, London WC1E 6BT, U.K.}
\ead{s.hogan@ucl.ac.uk}
\begin{abstract}
Beams of helium atoms in Rydberg-Stark states with principal quantum number $n=48$ and electric dipole moments of 4600~D have been decelerated from a mean initial longitudinal speed of 2000~m/s to zero velocity in the laboratory-fixed frame-of-reference in the continuously moving electric traps of a transmission-line decelerator. In this process accelerations up to $-1.3\times10^{7}$~m/s$^2$ were applied, and changes in kinetic energy of $\Delta E_{\mathrm{kin}}=1.3\times10^{-20}$~J ($\Delta E_{\mathrm{kin}}/e = 83$~meV) per atom were achieved. Guided and decelerated atoms, and those confined in stationary electrostatic traps, were detected \emph{in situ} by pulsed electric field ionisation. The results of numerical calculations of particle trajectories within the decelerator have been used to characterise the observed deceleration efficiencies, and aid in the interpretation of the experimental data.
\end{abstract}

\pacs{37.10.De, 32.80.Rm}

\section{Introduction}

Guides, decelerators and traps for gas-phase atoms and molecules which are based on two-dimensional arrays of surface-based metallic wires and electrodes~\cite{chiaverini05a,folman02a,meek08a,hogan12a} offer several advantages over larger three-dimensional structures. These chip-based devices (1)~permit the generation of a wide range of trap geometries, (2)~offer opportunities for miniaturisation using micro-fabrication techniques, (3)~can be readily extended, e.g., for the deceleration of heavier samples, (4)~are suited to operation in cryogenic environments, and (5)~can be integrated with other chip-based devices including, e.g., microwave circuits. In this context, a range of chip-based electric guides, decelerators and traps for atoms and molecules in highly-excited Rydberg states have recently been developed~\cite{hogan12a,allmendinger13a,lancuba13a,allmendinger14a,lancuba14a}. These devices exploit the large electric dipole moments associated with high Rydberg states, and the forces that can be exerted on samples in these states using inhomogeneous electric fields~\cite{wing80a,breeden81a,yamakita04a,vliegen04a}. With dipole moments exceeding 1000~D for states with principal quantum numbers, $n>16$, the large electric field gradients that can be generated above surface-based electrode structures permit efficient deceleration over short distances.

The chip-based Rydberg-Stark decelerators and traps that have been implemented up to now relied upon detection of decelerated or trapped atoms or molecules by pulsed electric field ionisation after exiting the devices. In experiments involving the production of velocity-controlled beams by acceleration, deceleration or guiding, samples were detected following a period of free-flight beyond the decelerators~\cite{hogan12a,allmendinger14a,lancuba14a}. Further, in experiments in which decelerated atoms were trapped on-chip in stationary electrostatic traps, it was necessary to reaccelerate the samples off the chip before detection~\cite{hogan12a,allmendinger13a}. This reduced the detection efficiency and complicated the interpretation of the rates of decay of the trapped samples. 

In experiments with Rydberg atoms and molecules confined in electrostatic traps composed of three-dimensional electrode structures, efficient \emph{in situ} detection of the trapped samples was achieved by pulsed electric field ionisation upon rapidly switching the potentials applied to the trap electrodes~\cite{vliegen07a,hogan08a,hogan09a,seiler11a}. Although detection of polar CO molecules, in the metastable a$\,^3\Pi_1$ state, by resonance-enhanced multiphoton ionisation directly above a chip-based Stark decelerator has been implemented~\cite{marx13a}, \emph{in situ} detection has so far not been possible in chip-based Rydberg-Stark decelerators. Here we demonstrate that efficient \emph{in situ} detection of samples of decelerated and electrostatically trapped helium Rydberg atoms can be achieved in a chip-based transmission-line decelerator~\cite{lancuba14a} following the application of pulsed electrical potentials to the ground-planes of the device. In the deceleration process atoms initially travelling at 2000~m/s were brought to rest in the laboratory-fixed frame-of-reference upon the application of accelerations of up to $-1.3\times10^{7}$~m/s$^2$. The mean kinetic energy removed per atom was therefore $\Delta E_{\mathrm{kin}}=1.3\times10^{-20}$~J ($\Delta E_{\mathrm{kin}}/e = 83$~meV).

In the following, the apparatus used in the experiments is first described in Section~\ref{sec:expt}. The experimental results are presented in Section~\ref{sec:results} and compared to the results of numerical calculations of particle trajectories within the transmission-line decelerator. Finally, the main aspects of the paper are summarised in Section~\ref{sec:concl} and comparisons are made with previous work in the literature. 

\begin{figure}
\begin{center}
\includegraphics[width = 0.85\textwidth]{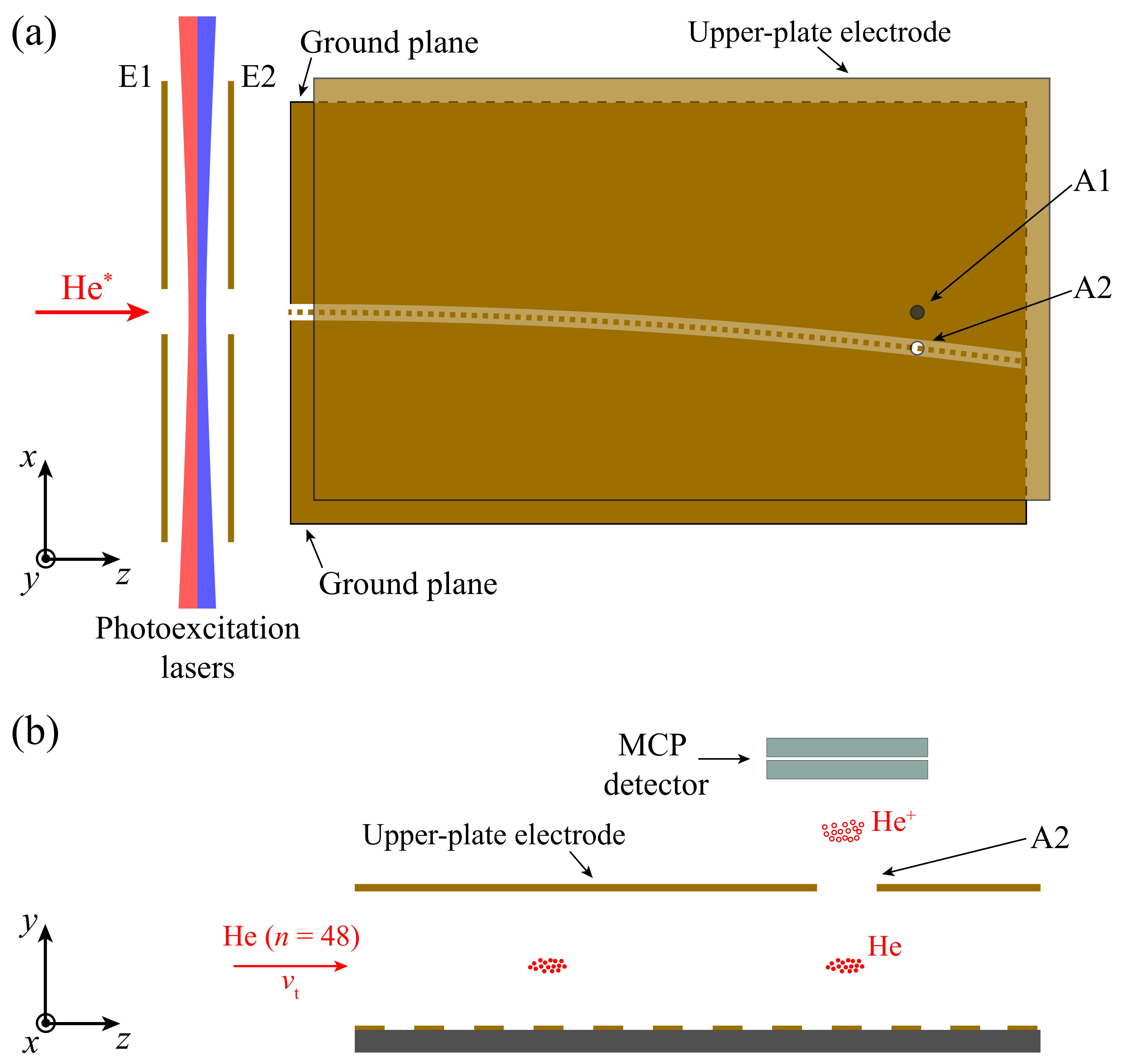}
\caption{Schematic diagram of the experimental apparatus (not to scale). (a) Top-view of the photoexcitation region and transmission-line decelerator in the $xz$-plane with the ion extraction apertures A1 and A2 in the upper-plate electrode indicated (see text for details). (b) Expanded side-view of the \emph{in-situ} detection region with the MCP detector located above the aperture A2 in the upper-plate electrode.}
\label{fig1}
\end{center}
\end{figure}

\section{Experiment}\label{sec:expt}

The experiments described here were performed using metastable helium atoms initially travelling in pulsed supersonic beams. These atoms, in the 1s2s\,$^3$S$_1$ level, were prepared in a dc electric discharge at the exit of a pulsed valve~\cite{halfmann00a}. After passing through a 2~mm diameter copper skimmer, and a region in which any stray ions or electrons travelling with the neutral atoms were deflected away from the propagation axis of the beam (see, e.g.~\cite{zhelyazkova15a}), the atoms entered the main part of the apparatus depicted schematically in Figure~\ref{fig1}. 

Between the two metallic plates labelled E1 and E2 in Figure~\ref{fig1} the metastable atoms were photoexcited to $n=48$ Rydberg-Stark states using the resonance-enhanced 1s2s\,$^3$S$_1\rightarrow$1s3p\,$^3$P$_2\rightarrow$1s$n$s/1s$n$d two-photon excitation scheme. This excitation process was driven using focussed cw laser radiation with wavenumbers of 25708.5876~cm$^{-1}$ ($\equiv388.9751$~nm) and 12698.4967~cm$^{-1}$ ($\equiv787.4948$~nm) for each transition, respectively. Photoexcitation was performed in a homogeneous electric field that was switched to $0.2$~V/cm for $3~\mu$s at the time of excitation. This was achieved by applying a pulsed potential of -0.25~V with rise and fall times of 300~ns, to electrode E2 while electrode E1 was maintained at 0~V. In this way $6$-mm-long bunches of Rydberg atoms were prepared in the low-field-seeking $|n,k\rangle=|48,25\rangle$ Rydberg-Stark state ($k$ is the difference between the parabolic quantum numbers $n_1$ and $n_2$~\cite{gallagher94a}). This state has a positive Stark energy shift and an electric dipole moment of 4600~D . States with $n=48$ were chosen for the experiments because (1)~the transition moments to the outer Stark states with the largest electric dipole moments were appropriate for efficient photoexcitation and deceleration, and (2)~they had a sufficiently high threshold for electric field ionisation that ionisation-induced losses during the deceleration process, particularly at the highest applied accelerations, were minimal. 
 
After photoexcitation the Rydberg atoms travelled toward the transmission-line decelerator [see Figure~\ref{fig1}(a)]. The general design of this decelerator was the same as that used in previous experiments~\cite{lancuba14a}. The center conductor of the device was composed of a set of equally spaced, square, copper segments, with dimensions of 1~mm $\times$ 1~mm, and a center-to-center spacing of $d_{\mathrm{cc}}=2$~mm. The insulating gap between the center-conductor segments and the decelerator ground planes was 1~mm and the upper-plate electrode was positioned parallel to, and 4.5~mm above, the resulting two-dimensional transmission line in the $y$-dimension. To reduce effects of collisions of decelerated and trapped Rydberg atoms with the trailing components of the atomic beam the decelerator was curved in the $xz$-plane with a radius of curvature of 0.81~m. The two-dimensional array of decelerator and ground-plane electrodes were prepared by chemical etching of copper on a FR4 laminate substrate.

To facilitate \emph{in situ} detection of the Rydberg atoms two 2.5-mm-diameter apertures were included in the upper-plate electrode of the decelerator. The first of these apertures [A1 in Figure~\ref{fig1}(a)] was located on the axis of propagation of the unperturbed atomic beam, while the second [A2 in Figure~\ref{fig1}(a)] was located 10~mm off axis in the negative $x$ dimension, directly above the 74$^{\mathrm{th}}$ segment of the decelerator center conductor. Rydberg atoms located within the decelerator structure at the positions of these two apertures were detected by electric field ionisation upon the simultaneous application of pulsed potentials of $+350$~V to the two ground planes of the transmission line. Immediately prior to the application of these ionisation pulses, the potentials on the upper-plate electrode and all electrodes of the decelerator center-conductor were switched to 0~V. The resulting He$^+$ ions were accelerated through the appropriate aperture in the upper-plate electrode and toward a microchannel plate (MCP) detector as depicted in Figure~\ref{fig1}(b). The MCP detector used had a front plate diameter of 25~mm which, with a potential of -2~kV applied, ensured that helium Rydberg atoms ionised beneath either aperture A1 or aperture A2 were efficiently collected for detection. When the decelerator was off, the excited Rydberg atoms propagated unperturbed through it and could be detected upon passing beneath aperture A1. If the decelerator was on, the strong inhomogeneous electric fields within it caused atoms which were not confined within one of the moving traps to be ionised or sufficiently deflected that they were unable to reach aperture A1. However, if transported through the decelerator while confined in one of the moving traps, the Rydberg atoms could be detected upon passing beneath aperture A2. 

The transmission-line decelerator was operated in the same manner as that described previously~\cite{lancuba14a}. Arrays of continuously moving electric traps were generated within the device by applying a set of five time-dependent, oscillating electrical potentials, $V_{i}(t)$, to consecutive segments, $i$, of the center conductor such that
\begin{eqnarray}
V_{i}(t) &=& V_0\cos[\omega t - (i-1)\phi],
\end{eqnarray}
where $\omega$ is the angular oscillation frequency of the potentials, and $\phi=2\pi/5$ is the phase shift from one center-conductor segment to the next. For all experiments described here the amplitude of the oscillating potentials was selected to be $V_0=150$~V, and the time-independent potential on the upper-plate electrode, $V_{\mathrm{u}}$, was set to $V_{\mathrm{u}}=-V_0/2$. With this configuration of electrical potentials the tangential speed, $v_{\mathrm{t}}$, at which the traps moved through the curved decelerator was 
\begin{eqnarray}
v_{\mathrm{t}} = \frac{5d_{\mathrm{cc}}\omega}{2\pi},
\end{eqnarray}
and could be controlled by adjusting $\omega$. To accelerate or decelerate the traps, and in doing so accelerate or decelerate the bunches of atoms confined within them, a time-dependence was introduced to this angular oscillation frequency. For a selected initial speed $v_{\mathrm{t}}(0)$, and a constant tangential acceleration $a_{\mathrm{t}}$, the resulting time-dependent angular frequency, $\omega(t)$, was
\begin{eqnarray}
\omega(t) &=& \omega(0) + \frac{\pi a_{\mathrm{t}}}{5\,d_{\mathrm{cc}}}t,
\end{eqnarray}
where $\omega(0)=2\pi v_{\mathrm{t}}(0)/(5\,d_{\mathrm{cc}})$. The time-dependent potentials used in the experiments were generated at low voltage using a set of arbitrary waveform generators and amplified by a factor of~50 before being applied to the decelerator electrodes.

\begin{figure}
\begin{center}
\includegraphics[width = 0.65\textwidth]{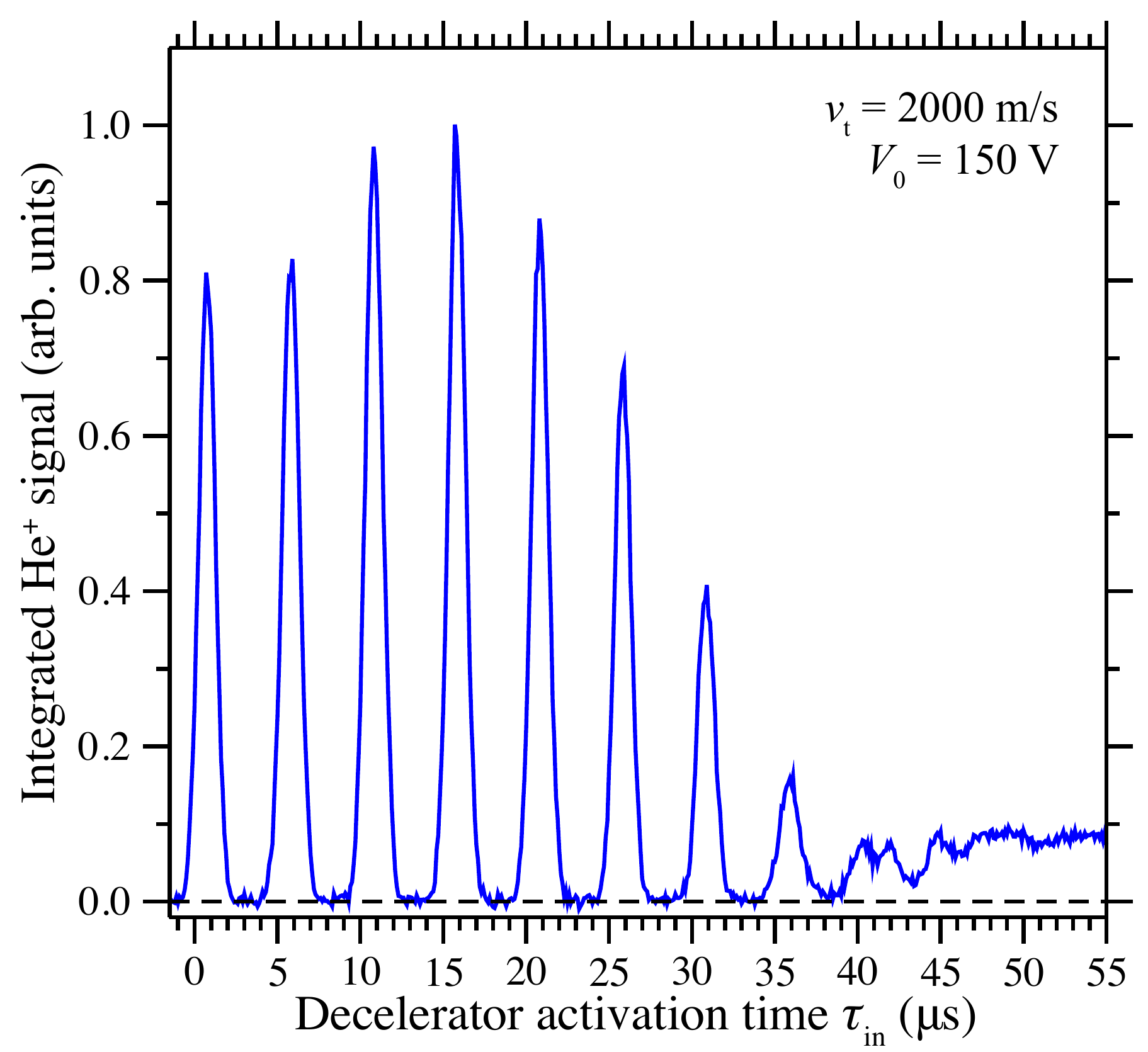}
\caption{Dependence of the measured signal from helium Rydberg atoms guided at a constant speed of 2000~m/s while confined in the travelling electric traps of the transmission-line decelerator on the time-delay, $\tau_{\mathrm{in}}$, between photoexcitation and activation of the decelerator potentials. The measurement was made for a fixed flight time from the photoexcitation region to the detection position beneath aperture A2 of $80.5~\mu$s, and with $V_0=150$~V.}
\label{fig2}
\end{center}
\end{figure}

\section{Results}\label{sec:results}

The operation of this transmission-line decelerator with \emph{in situ} detection was investigated by measuring the times of flight of the helium Rydberg atoms from their position of photoexcitation to the detection apertures A1 or A2 [see Figure~\ref{fig1}(a)]. For the bunches of Rydberg atoms with mean initial longitudinal speeds of 2000~m/s that were prepared in the experiments, the intensity maxima of the corresponding time-of-flight distributions were recorded 80.5~$\mu$s after photoexcitation when the device was off.

To determine the optimal time after photoexcitation to activate the oscillating decelerator potentials, $\tau_{\mathrm{in}}$, the detection time of flight was set to 80.5~$\mu$s, and the signal associated with atoms guided through the decelerator at a constant speed of 2000~m/s, while confined in a single travelling trap, was recorded while the activation time of the decelerator was adjusted. The results of this measurement are presented in Figure~\ref{fig2}. From this data is can be seen that intensity maxima appear in the signal from the guided atoms every time the decelerator potentials were activated when the bunch of Rydberg atoms was located at the position of one of the moving traps of the device. Because the detected atoms initially moved at 2000~m/s, and the spatial separation between the traps of the decelerator was $5\,d_{\mathrm{cc}}=10$~mm, the intensity maxima seen in Figure~\ref{fig2} are separated by the time taken by the atoms to move from the position of one trap to the next, i.e., $5~\mu$s. 

At early activation times, i.e., $\tau_{\mathrm{in}}\simeq1~\mu$s, the oscillating decelerator potentials are active before the Rydberg atoms exit the photoexcitation region through the aperture in E2 [see Figure~\ref{fig1}(a)], and are loaded into the first trap of the decelerator as it forms at the entrance of the device. For $\tau_{\mathrm{in}}>5.7~\mu$s the atoms are located within the decelerator when the device is activated. The contrast in the signal, between the intensity maxima and minima, at early times in Figure~\ref{fig2} is evidence that Rydberg atoms that are not transported through the decelerator while confined in one of the travelling traps are either ionised or deflected sufficiently that they do not reach either of the detection apertures at the selected flight time. When a long time elapses between photoexcitation and activation of the decelerator potentials, i.e., when $\tau_{\mathrm{in}}\gtrsim 40~\mu$s, the atoms are no-longer located above the curved center-conductor of the decelerator when the traps are generated. As a result, they are not efficiently loaded into the travelling traps and guided to the detection position beneath aperture A2. In this situation some atoms do reach aperture A1 at the appropriate time of flight for detection but because they are not guided while confined in one of the travelling traps, the observed signal intensity does not depend upon the activation time of the decelerator potentials. To maximise the length of the decelerator used for deceleration of these fast beams of helium, in all of the measurements described below a decelerator activation time of $\tau_{\mathrm{in}}=0.7~\mu$s was employed.

\begin{figure}
\begin{center}
\includegraphics[width = 0.99\textwidth]{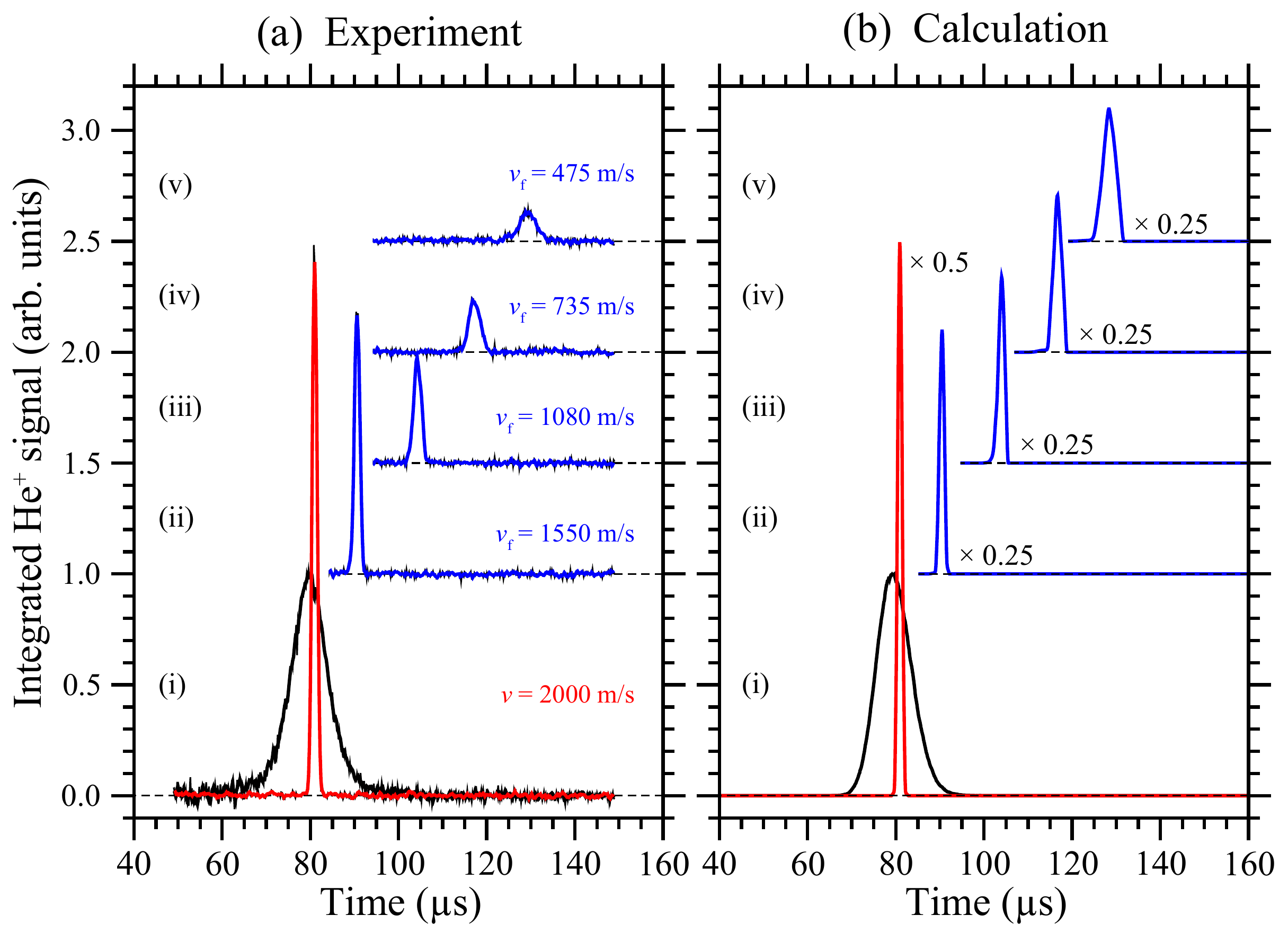}
\caption{(a) Measured and (b) calculated helium Rydberg atom time-of-flight distributions with (i)~the decelerator off (broad black curve), and operated to guide atoms at a constant speed of 2000~m/s (narrow red curve), (ii)~$a_{\mathrm{t}}=-0.5\times10^{7}$~m/s$^2$ and $v_{\mathrm{f}}=1550$~m/s, (iii)~$a_{\mathrm{t}}=-0.9\times10^{7}$~m/s$^2$ and $v_{\mathrm{f}}=1080$~m/s, (iv)~$a_{\mathrm{t}}=-1.1\times10^{7}$~m/s$^2$ and $v_{\mathrm{f}}=735$~m/s, and (v)~$a_{\mathrm{t}}=-1.2\times10^{7}$~m/s$^2$ and $v_{\mathrm{f}}=475$~m/s. In each case $V_0=150$~V. For clarity of presentation the upper sets of data in (a) and (b) are vertically offset.}
\label{fig3}
\end{center}
\end{figure}

The effect of the decelerator on the time-of-flight distributions of the Rydberg atoms was investigated by recording the integrated He$^+$ signal measured for a range of times of flight from the excitation position to the \emph{in situ} detection regions. With the decelerator off, the time-of-flight distribution associated with the unperturbed beam of Rydberg atoms was recorded and is displayed in Figure~\ref{fig3}(a-i) (broad black curve). With the decelerator then activated to guide the trapped atoms at a constant speed of 2000~m/s to the detection region beneath aperture A2, the narrower and more intense time-of-flight distribution also displayed in Figure~\ref{fig3}(a-i) was recorded. The single intensity maximum associated with this distribution confirms that the Rydberg atoms were all transported through the decelerator in one single travelling trap. The reduced width and increased intensity of the signal, compared to that recorded with the decelerator off, is a consequence of the strong confinement of the atoms within the trap as they were guided.

The trapped atoms transported through the decelerator at constant speed in Figure~\ref{fig3}(a-i) were then decelerated by applying the appropriate frequency chirps to the oscillating decelerator potentials. For a constant acceleration of $a_{\mathrm{t}}=-0.5\times10^{7}$~m/s$^2$ the atoms were decelerated from 2000~m/s to 1550~m/s and arrived at the detection region beneath aperture A2 after a time-of-flight of $90.5~\mu$s [Figure~\ref{fig3}(a-ii)]. The reduction in the intensity of the signal associated with these decelerated atoms is a consequence of the effect of the acceleration on the trapping potential experienced by the atoms, and collisions involving the trapped atoms. Larger accelerations, of $a_{\mathrm{t}}=-0.9\times10^{7},~ -1.1\times10^{7}$ and $-1.2\times10^{7}$~m/s$^2$ [Figure~\ref{fig3}(a-iii) -- Figure~\ref{fig3}(a-v) as indicated], resulted in the production of velocity-controlled beams with final tangential velocities of $1080,~ 735$ and $475$~m/s, respectively.

Comparison of the experimentally recorded Rydberg atom time-of-flight distributions presented in Figure~\ref{fig3}(a) with the results of numerical calculations of particle trajectories through the decelerator [Figure~\ref{fig3}(b)] offers insight into the origins of the loss of atoms upon deceleration to lower final velocities. These calculations were performed by solving the classical equations of motion of the Rydberg atoms in the moving frame of reference associated with one single trap of the decelerator. The effects of acceleration in this reference frame were included through pseudo potentials which were dependent upon the displacement from the electric field minimum of the trap~\cite{lancuba14a,meek09a,hogan13a}. Throughout the entire guiding and deceleration processes the instantaneous centripetal acceleration, arising as a result of the curvature of the decelerator, and the applied tangential acceleration, $a_{\mathrm{t}}$, were accounted for~\cite{lancuba14a}.

Because the \emph{in situ} detection regions in the decelerator have very well defined spatial dimensions, the longitudinal temperature of the excited ensemble of Rydberg atoms could be accurately determined from the time-of-flight distribution recorded with the decelerator off. Using this data the bunch of atoms was found to have a relative translational temperature in the $z$~dimension of $T_{z}=\langle E_{\mathrm{kin}}\rangle/k_{\mathrm{B}}=3$~K,  where $\langle E_{\mathrm{kin}}\rangle$ is the mean relative kinetic energy of the atoms, and $k_{\mathrm{B}}$ is the Boltzmann constant. Taking into account the 6~mm longitudinal extent of the excited bunch of atoms, this temperature matches that reported in other experiments with pulsed supersonic beams of metastable helium generated with similar discharge sources~\cite{allmendinger13a}. The transverse temperatures  of the atoms were determined, from the geometric constraints imposed by the skimmer and focussed narrow-bandwidth laser beams in the experiments, to be $T_{x}=T_{y}=10$~mK. By randomly generating ensembles of Rydberg atoms with these initial parameters, trajectories in the moving traps of the decelerator were calculated numerically until the traps reached the detection aperture A2. The corresponding arrival times of the atoms in this detection region were then used to generate the calculated time-of-flight distributions presented in Figure~\ref{fig3}(b). 

In Figure~\ref{fig3}(b-i) the calculated intensity of the signal associated with the Rydberg atom beam when the decelerator was off, was normalised in the same way as that recorded experimentally [Figure~\ref{fig3}(a-i)]. The result of the particle trajectory calculations performed for atoms guided at a constant speed of 2000~m/s exhibits an intense, narrow maximum in the time of flight distribution at a flight time of $80.5~\mu$s. This time-of-flight distribution is in agreement with that observed in the experimental data if it is scaled by a factor of 0.5. This scaling accounts for the differences in the He$^+$ ion extraction efficiency from the \emph{in situ} detection region beneath aperture A1, and that in the region beneath aperture A2 in the experiment. It also accounts for effects of collisions between the trapped Rydberg atoms on the efficiency with which they are transported through the decelerator~\cite{lancuba14a}. Neither of these effects are treated in the calculations. 

For a tangential acceleration of $a_{\mathrm{t}}=-0.5\times10^{7}$~m/s$^2$ the intensity maximum in the corresponding calculated time-of-flight distribution [Figure~\ref{fig3}(b-ii)] shifts to the later time of $90.5~\mu$s, exactly matching that observed in the experiments. The calculated flight times associated with the intensity maxima upon further deceleration, toward final speeds of 475~m/s, are also in excellent agreement with those in the experimental data. However, in each case the amplitude of the intensity maxima in the time of flight distributions [scaled by a factor of 0.25 with respect to the undecelerated distribution as indicated in Figure~\ref{fig3}(b)] is greater in the calculated data than in the experimental data. The differences in the signal intensity indicate a loss of atoms from the travelling traps of the decelerator in the experiments which is not solely a result of their translational motion. This decay of atoms from the moving traps of the decelerator arises from a combination of spontaneous emission, effects of blackbody radiation, collisions between the Rydberg atoms within the traps, and collisions between the trapped Rydberg atoms and the background gas in the vacuum chamber and is discussed in more detail in Section~\ref{sec:concl}.

\begin{figure}
\begin{center}
\includegraphics[width = 0.6\textwidth]{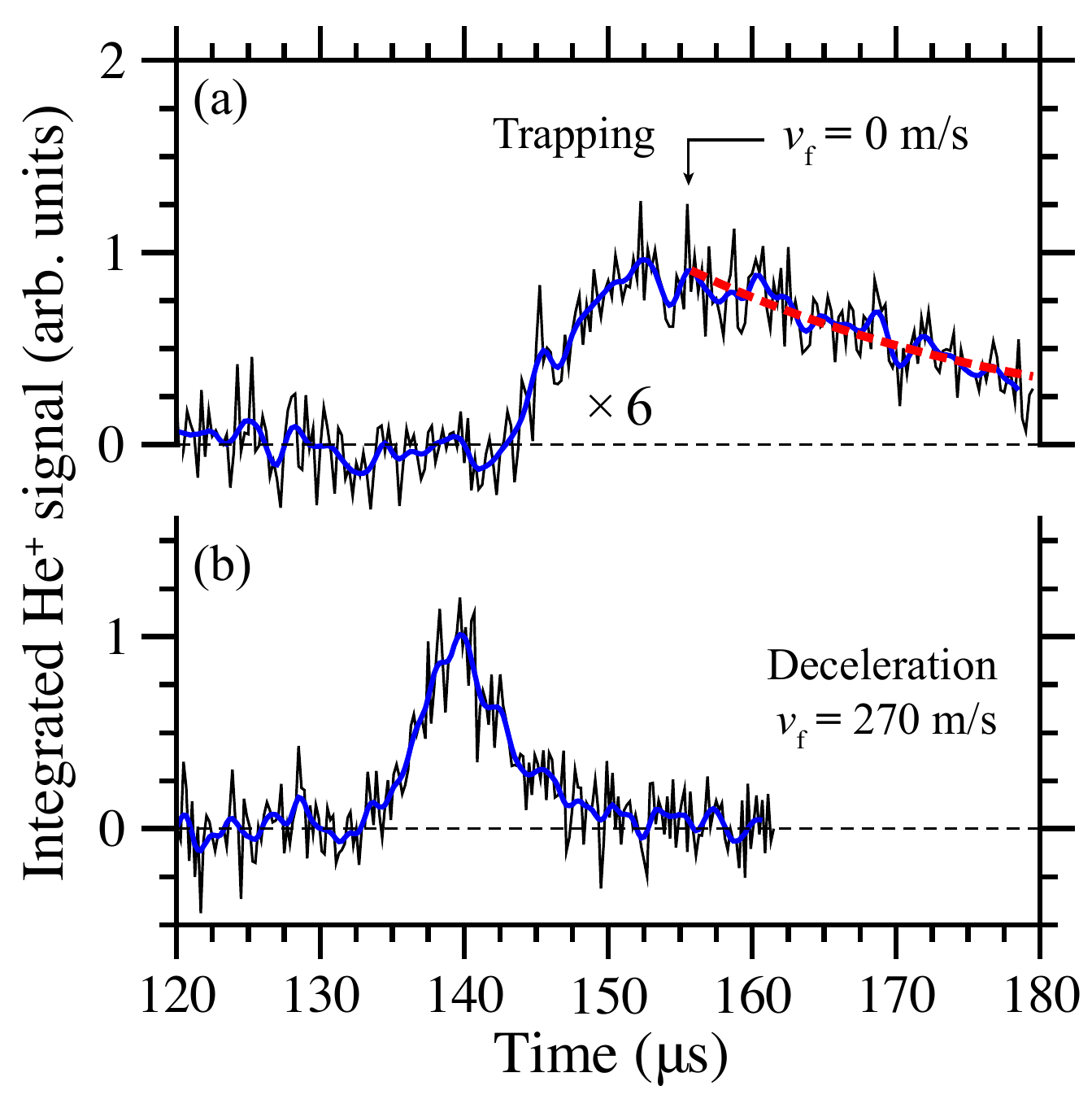}
\caption{Helium Rydberg atom time-of-flight distributions for (a) deceleration from 2000~m/s to zero-velocity in the laboratory-fixed frame of reference with $a_{\mathrm{t}}=-1.279\times10^{7}$~m/s$^2$, and (b) deceleration to $v_{\mathrm{f}}=270$~m/s with $a_{\mathrm{t}}=-1.250\times10^{7}$~m/s$^2$. In (a) the electric traps of the decelerator are stationary and located directly beneath aperture A2 for all times beyond 156~$\mu$s, indicated by the arrow, and the data has been scaled by a factor of 6 with respect to (b) for clarity of presentation. For each measurement $V_0=150$~V.}
\label{fig4}
\end{center}
\end{figure}

In the case of the most slowly moving beams of Rydberg atoms, decelerated to 475~m/s, in Figure~\ref{fig3}, $94\%$ of the initial kinetic energy was removed for $a_{\mathrm{t}}=-1.2\times10^{7}$~m/s$^2$. Therefore with a further increase in the acceleration the trapped atoms could be brought to a stand-still in the laboratory-fixed frame of reference. As can be seen in Figure~\ref{fig4}(b), increasing the acceleration to $a_{\mathrm{t}}=-1.250\times10^{7}$~m/s$^2$ results in a final tangential velocity of 270~m/s. In this case the decelerated atoms pass beneath aperture A2 $139~\mu$s after photoexcitation. A further adjustment of the acceleration to $a_{\mathrm{t}}=-1.279\times10^{7}$~m/s$^2$ brought the trap to rest directly beneath aperture A2, $156~\mu$s after photoexcitation [vertical arrow in Figure~\ref{fig4}(a)]. The data in this figure have been scaled by a factor of 6, highlighting the further loss of trapped atoms over the final period of deceleration. 

When the trap in which the Rydberg atoms are decelerated is brought to rest directly beneath aperture A2, the changes observed in the signal intensity in the time-of-flight distribution, i.e., that for times beyond $156~\mu$s, represent an \emph{in situ} measurement of the decay of the atoms from the stationary electrostatic trap. In the absence of collisions and blackbody radiation, the decay rates via spontaneous emission from the initially excited $n=48$ Rydberg-Stark states are on the order of 1000~s$^{-1}$ (i.e, $\tau_{1/e}\sim1$~ms). However, fitting a single exponential function to the data in Figure~\ref{fig4}(a) for times beyond $156~\mu$s results in a decay constant of only $25~\mu$s (dashed red curve). Because the experiments were performed in a room temperature environment, contributions from blackbody radiation to the evolution and decay of the trapped atoms will be significant~\cite{seiler11a,seiler12a}. These effects, particularly those associated with blackbody photoionisation, and collisions of the trapped Rydberg atoms with the background gas dominate the observed decay from the trap.

\section{Discussion and conclusions}\label{sec:concl}

The time constant of $25~\mu$s associated with the decay of the Rydberg atoms from the stationary off-axis electrostatic trap employed in the experiments described here, although significantly shorter than the fluorescence lifetimes of the excited states, is of a similar magnitude to the time constants reported in previous experiments in which Rydberg atoms were confined in stationary chip-based traps constructed from similar materials. For example, in experiments with hydrogen atoms in Rydberg-Stark states with $n=31$ trapped on the atomic beam axis above a surface-electrode decelerator, decay constants of $\tau_{1/e}\simeq30~\mu$s were determined~\cite{hogan12a}. While in similar experiments with helium atoms in singlet Rydberg states with $n=31$, trap decay constants of $\tau_{1/e}=22.7~\mu$s were reported~\cite{allmendinger13a}. In both cases, these decay constants were shorter than the respective, $142~\mu$s and $58~\mu$s, fluorescence lifetimes of the excited states. Because the atoms were trapped on the axes of the atomic beams in these previous experiments, collisions with the trailing components of the beams must have played a role in the decay from the traps~\cite{seiler11a}. In the experiments reported here, the Rydberg atoms were trapped off the axis of propagation of the atomic beam, in a transmission-line decelerator of a different design to the surface-electrode decelerators. However, the materials from which the chip-based array of decelerator electrodes were fabricated are similar (chemically etched copper on a FR4 laminate substrate). Because these laminate substrates outgas in vacuum, we suspect that collisions between the trapped Rydberg atoms and the gas emanating from the substrates used in the construction of the decelerators play a significant role in the observed trap decay. 

\begin{figure}
\begin{center}
\includegraphics[width = 0.65\textwidth]{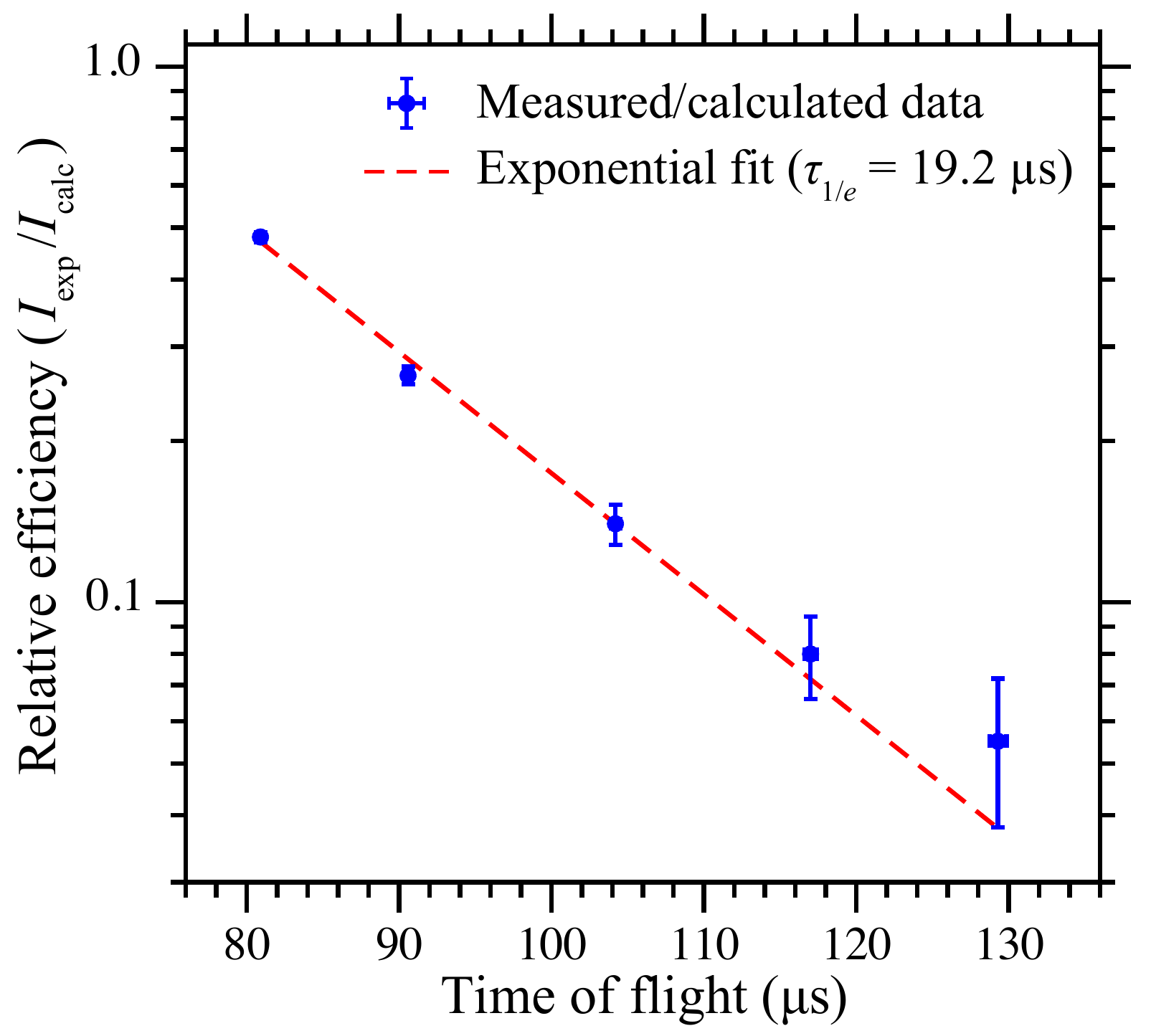}
\caption{Ratios of the peak signal intensities in the experimentally recorded time-of-flight distributions in Figure~\ref{fig3}(a) to those determined in the numerical particle trajectory calculations in Figure~\ref{fig3}(b). The time of flight associated with each intensity maximum is indicated on the horizontal axis. The dashed red curve represents the result of fitting a single exponential function with a time constant of $\tau_{1/e}=19~\mu$s to the set of data points.}
\label{fig5}
\end{center}
\end{figure}

Further quantitative information on the rates of decay of atoms confined in the travelling traps of the decelerator during the deceleration process, can be obtained by comparing the experimentally recorded and calculated Rydberg atom time-of-flight distributions presented in Figure~\ref{fig3}. This was achieved by determining the ratio of the peak intensity in each experimentally recorded time-of-flight distribution in this figure, $I_{\mathrm{exp}}$, to that in the corresponding calculated time-of-flight distribution, $I_{\mathrm{calc}}$. The dependence of these relative efficiencies on the flight time of the atoms is displayed in Figure~\ref{fig5} (shaded circles). In calculating these relative signal intensities, the dependence of the deceleration efficiency on the applied acceleration is deconvolved, with the result that the rate of decay of atoms from the travelling decelerator traps is obtained. As can be seen in Figure~\ref{fig5}, to first order these data exhibit an exponential decay with a time constant of $19.2~\mu$s. This is again on the same scale as the time constant associated with the decay of the Rydberg atoms from the stationary trap [Figure~\ref{fig4}(a)]. These results, together with the observations from previously reported experiments, would therefore suggest that in future experiments with chip-based Rydberg-Stark decelerators it will be important to study the effects of the substrate materials on the rates of decay of trapped atoms and molecules, and possibly refine the construction of the devices using alternative substrate materials based on the outcomes of this work. 

In conclusion, we have demonstrated complete deceleration and \emph{in situ} detection of fast beams of helium Rydberg atoms to zero-velocity in the laboratory-fixed frame of reference while confined in the travelling electric traps of a transmission-line decelerator. When brought to rest the samples of decelerated Rydberg atoms were electrostatically trapped directly above the center conductor of the segmented electrical transmission line used to generate the deceleration and trapping electric field distributions. These results are expected to be of importance in collision and spectroscopic studies involving cold Rydberg atoms and molecules, including those at vacuum--solid-state interfaces~\cite{hill00a,thiele14a,gibbard15a}, and in hybrid cavity quantum electrodynamics with Rydberg atoms coupled to chip-based microwave circuits~\cite{hogan12b,carter12a}. Since the segmented electrical transmission line of the decelerator used here can be considered to be composed of a series of coplanar electrical resonators, the results presented in Figure~\ref{fig4}(a) represent a demonstration of electrostatic trapping gas-phase Rydberg atoms above such a circuit element. 

\ack

This work was supported financially by the Department of Physics and Astronomy and the Faculty of Mathematical and Physical Sciences at University College London (UCL), the UCL Impact Studentship Program, and the Engineering and Physical Sciences Research Council (EPSRC) under Grant No. EP/L019620/1.

\end{document}